\RequirePackage{fix-cm}
\documentclass[12pt]{amsart}       
\usepackage{graphicx}
\usepackage{mathptmx}      
\usepackage{latexsym}
\usepackage{color}
\usepackage{mathptmx}
\usepackage{url}
\usepackage{bm}
\usepackage{cite}
\usepackage{amsmath}
\usepackage{amssymb}
\usepackage{bbm}
\usepackage{graphicx}
\usepackage{subfigure}
\usepackage{tikz}
\usepackage{array}
\usepackage{multirow, makecell}
\usetikzlibrary{arrows,positioning,shapes.geometric}
\usetikzlibrary{matrix,chains,decorations.pathreplacing}
\usepackage{xcolor}
\usepackage{microtype}
\usepackage{booktabs}
\usepackage{url,diagbox}
\newcommand{\ud}{\,\mathrm{d}}

\begin{document}

\title{Recycling cardiogenic artifacts in impedance pneumography}
\author{Yao Lu\textsuperscript{1} \and Hau-tieng Wu\textsuperscript{2,3} \and John Malik\textsuperscript{2}}
\thanks{\textsuperscript{1} Department of Laboratory Medicine and Pathobiology, University of Toronto, Toronto, ON, Canada}
\thanks{\textsuperscript{2} Department of Mathematics, Duke University, Durham, NC, USA}
\thanks{\textsuperscript{3} Department of Statistical Science, Duke University, Durham, NC, USA}





\maketitle

\begin{abstract}
\textit{Purpose:} 
Biomedical sensors often exhibit cardiogenic artifacts which, while distorting the signal of interest, carry useful hemodynamic information. 
We propose an algorithm to remove and extract hemodynamic information from these cardiogenic artifacts.
\textit{Methods:} We apply a nonlinear time-frequency analysis technique, the de-shape synchrosqueezing transform {(dsSST)}, to adaptively isolate the high- and low-frequency components of a single-channel signal.
We demonstrate this technique's effectiveness by {removing and} deriving hemodynamic information from the cardiogenic artifact in an impedance pneumography (IP). 
\textit{Results:} The instantaneous heart rate is extracted, and the cardiac and respiratory signals are reconstructed. 
\textit{Conclusions:} The {dsSST} is suitable for generating useful {hemodynamic} information from the cardiogenic artifact in a single-channel IP.
We propose that the usefulness of the dsSST as a recycling tool extends to other biomedical sensors exhibiting cardiogenic artifacts.
\end{abstract}

\section{Introduction}

The popularity and quality of physiological measurement devices have grown significantly in recent years \cite{STOREY201750}. These devices see increasing applicability because of fields like mobile health \cite{piwek2016rise, electronics3020282}. In critical clinical scenarios like the intensive care unit or the operating room, we are able to afford multiple sensors, each one optimized to monitor a specific physiological system.  A typical example is the patient monitor commonly seen at the bedside.
However, this is not always the case. For example, in an ambulance, only a few sensors are available, and for most mobile health applications, only one or two sensors are used. Even in environments where patient monitors are present, due to technical problems, we may not be able to trust the quality of the information recorded. For example, when the respiratory flow channel is dead, we can count on the respiratory information hidden in the photoplethysmogram (PPG).
Therefore, every bit of information is precious in any scenario.

We want to maximize the quantity of clinical information extracted from physiological signals. The main challenge is separating information which has been mixed together in one channel. 
For example, the respiratory information exists as the amplitude modulation of the electrocardiogram (ECG), while it exists as the low-frequency component of the PPG signal. Furthermore, the cardiac information exists as the high-frequency component of the respiratory flow signal, etc. 
In general, when this mixing of signals appears as the sum of multiple oscillatory components, the task of separating them could be understood as the {\em single channel blind source separation (scBSS)} problem. Under some conditions, the problem can be easily solved by applying a bandpass filter. When extra channels are available, an adaptive filtering scheme can be helpful. A more sophisticated approach is required when only one channel is available. 

The scBSS problem is more complicated when we enter the field of mobile health because of the presence of artifacts.
The term ``artifact'' refers to an element of the recording that is irrelevant to the information in which we have interest. Usually, artifacts are not modeled as random noise but rather as a signal with some structure. 
When the artifact's structure follows a set of rules, these rules can help us remove the artifact. If the rules are physiological, we may even turn the artifact into useful physiological information, and this is our target in this paper. Typical examples include the cardiogenic artifact in the respiratory signal or the respiratory artifact in the PPG signal. 
We focus on the scBSS problem when there is only a single recorded channel available. We use a nonlinear-type time-frequency filtering strategy to extract as much information as possible from a single-channel recording.

\subsection{Cardiogenic artifacts}
Take the commonly seen cardiogenic artifact in biomedical measurements as an example of maximizing information quantity. 
Cardiogenic artifacts 
come from the blood volume movement in the thorax \cite{BrownBarber1994} and
may mask the physiological information in which we have interest. Examples of measurements in which cardiogenic artifacts may be observed include the impedance pneumography (IP) \cite{Weese-Mayer1989,Luo1994,Folke2003},
the respiratory inductive plethysmography (RIP) \cite{Zhang2012}, thoracic or abdominal movements recorded via piezo-electric band \cite{Lin_Wu_Hsu_Wang_Huang_Huang_Lo:2016}, 
the capnogram \cite{Smith1994}, the electroencephalogram \cite{1741-2552-12-3-031001} and the
esophageal pressure signal \cite{Schuessler1998}. 
Published algorithms intended for clinical deployment have focused on removing cardiogenic artifacts using techniques such as digital filtering \cite{Wilson1982,Poupard2008}, adaptive filtering \cite{Luo1994,Yasuda2005}, template subtraction via the electrocardiogram \cite{Schuessler1998,Seppa2011}, the synchrosqueezing transform \cite{capnogram:2018}, or blind source separation \cite{1741-2552-12-3-031001}. 
Removing the cardiogenic artifact enhances and clarifies the information relayed to medical professionals; it accurately estimates the main information present in the recording and allows for its prudent utilization. 

Filtering out the cardiogenic artifact to enhance the main signal is a strategy suitable for clinical deployment because of the variety of resources available in such settings; hemodynamic information may be measured using standard methods such as the electrocardiogram.
However, when only a few channels are available, the cardiogenic artifact is valuable because it provides additional physiological information. Utilization of the cardiogenic artifact has been suggested and applied several times in the past.
For example, in a thoracocardiography \cite{Sackner1991613,Bucklar2003}, cardiac output is estimated by first separating the cardiac waveforms from the dominant respiratory signal.
In \cite{Lin_Wu_Hsu_Wang_Huang_Huang_Lo:2016}, the authors suggest using the cardiogenic artifact in thoracic and abdominal movements via piezo-electric band to detect central sleep apnea. 

\subsection{Our contributions}
In this paper, we apply two recently developed nonlinear-type time-frequency (TF) analysis techniques, namely the synchrosqueezing transform (SST) and the de-shape SST (dsSST), to achieve the scBSS task and hence maximize the quantity of physiological information which may be extracted from an individual signal.  We demonstrate how to estimate instantaneous heart rate (IHR) from the cardiogenic artifact in a single-channel IP, and we use this estimate to separate the IP into its respiratory and hemodynamic components. The algorithm is applied to nineteen patients undergoing bronco-endoscopies for the sake of diagnosing pulmonary disease.
The MATLAB code is made publicly available so that our methods may be reproduced and applied to other signals for other purposes.

\section{Modeling the cardiogenic artifact}\label{Sect:2}

We model the cardiogenic artifact by the wave-shape function and the {\em adaptive non-harmonic model} \cite{Wu:2013,lin2016waveshape}.
The adaptive non-harmonic model is motivated by a need to describe oscillatory physiological phenomena such as respiration and cardiac activity. 
There are several facts about oscillatory physiological signals that interest us. The amplitude of the oscillation, the cycle period, and the oscillation pattern may vary from one cycle to another. Moreover, one physiological signal may not contain only one oscillatory component. Take the IP into account, whose primary purpose is to measure respiration. The amplitude becomes larger and the cycle period becomes longer when taking a deep breath, and the oscillatory pattern might change when the inhalation and exhalation pattern changes. In addition to the respiratory oscillation in the IP, the cardiac activity is also recorded as another oscillatory component. Since this recorded cardiac activity is not the main purpose of the IP, it is commonly viewed as nuisance and called the cardiogenic artifact. Due to heart rate variability, the cardiac activity also has a time-dependent period and might have a time-varying amplitude and oscillatory pattern.

\begin{figure*}
\centering
 $\hspace{-0.7in} \vcenter{\includegraphics[width=0.8\textwidth]{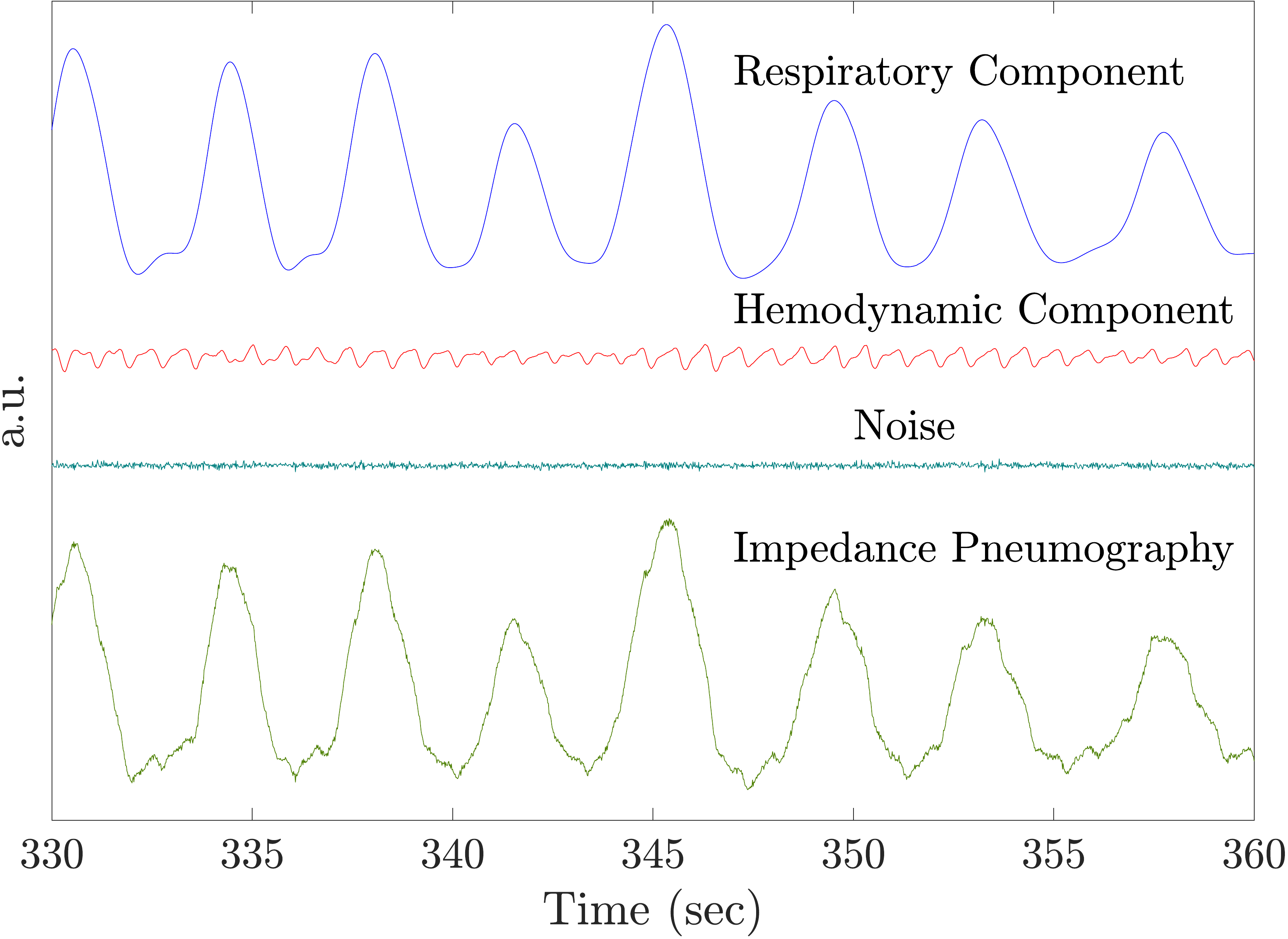}}\hspace{-2.2in}\vcenter{\includegraphics[width=0.2\textwidth]{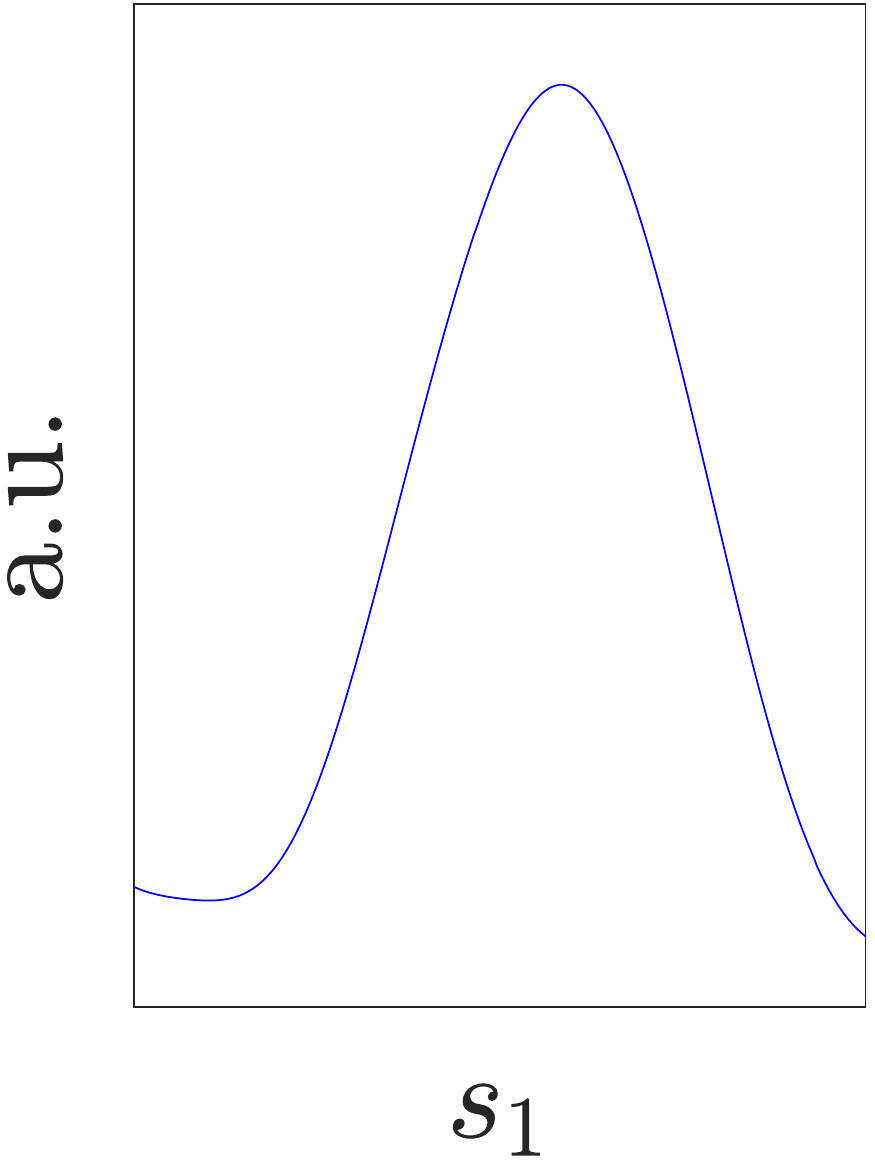} \\
 \vspace{0.1in} \includegraphics[width=0.2\textwidth]{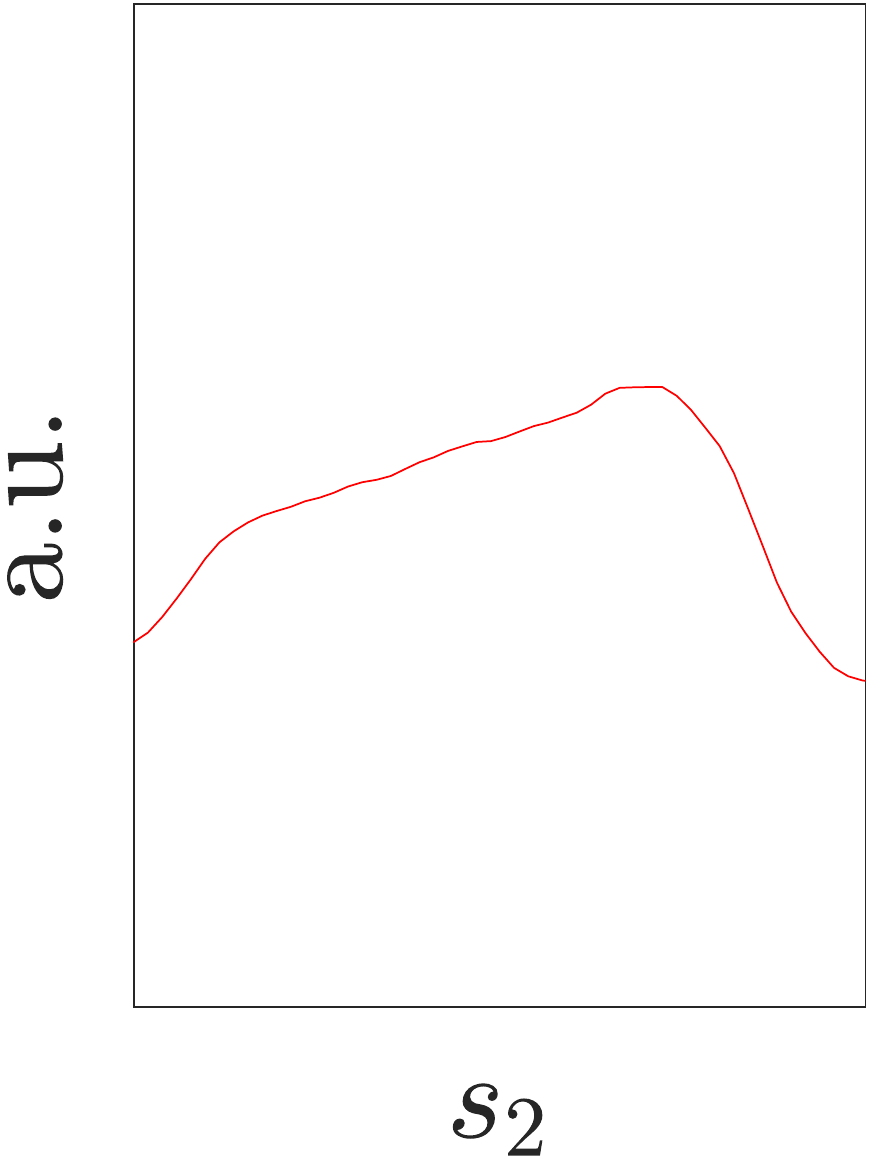}\vspace{0.16in}}$
\caption{An illustration of modeling the IP by the adaptive non-harmonic model. We plot (from top to bottom) the respiratory component, the hemodynamic component, generated noise, and the recorded IP. We highlight the respiratory wave-shape function ($s_1$) and the cardiac wave-shape function ($s_2$) on the right-hand side.}\label{fig1111}
\end{figure*}

Clearly, these signals cannot be modeled using simple harmonic functions or without regard to the time-varying nature of their morphologies. In \cite{Wu:2013}, the following {\em adaptive non-harmonic model} is proposed to capture signals of this kind:
\begin{equation}\label{Model:equation}
f(t) = \sum_{l=1}^L a_l(t)s_l(\phi_l(t))+\Phi(t),
\end{equation}
where $f$ is the recorded signal, $L$ describes the number of oscillatory components, $a_l(t)$ describes the amplitude of the $l$\textsuperscript{th} oscillatory component (which is assumed to be positive), $s_l$ describes the oscillatory pattern of the $l$\textsuperscript{th} component (which is assumed to be $1$-periodic), $\phi_l$ describes the phase of the $l$\textsuperscript{th} component (which is assumed to be monotonically increasing so that $1/\phi'_l(t)$ describes the period of the cycle at time $t$), and $\Phi$ is assumed to be the noise that contaminates the signal. 
We call $a_l(t)$ the amplitude modulation (AM) of the $l$\textsuperscript{th} component. Note that the inverse of the period of a cycle is in general understood as the frequency, so $\phi_l'(t)$ is called the {\em instantaneous frequency (IF)} of the $l$\textsuperscript{th} component. Finally, $s_l$ is called the {\em wave-shape function}, which captures the oscillatory pattern of the $l$\textsuperscript{th} signal. The model is called ``non-harmonic'' since the oscillation is non-sinusoidal. For the IP, $L=2$ since it contains not only the respiratory signal, but also the cardiogenic artifact. See Fig.~\ref{fig1111} for an example.

This model is also considered in several other works, for example, \cite{IMCStefanovska:2015,HouShi:2016,XuYangDaubechies:2018}. Mild assumptions are needed for the adaptive non-harmonic model to be well-behaved. The model in (\ref{Model:equation}) is further expanded in \cite{lin2016waveshape} to capture the time-varying oscillatory pattern. The generalized model described in \cite{lin2016waveshape} is more complicated, but its essence is the same as that in (\ref{Model:equation}), and (\ref{Model:equation}) is enough for the purpose of this work. Therefore, to keep the discussion simple, we satisfy ourselves with (\ref{Model:equation}) to avoid distracting the focus. We refer readers with interest to \cite{lin2016waveshape} for technical details.

\section{The de-shape synchrosqueezing transform}

Motivated by biomedical signals and the adaptive non-harmonic model (\ref{Model:equation}), the dsSST is designed in \cite{lin2016waveshape} as a signal processing tool to extract information from $f$ defined in (\ref{Model:equation}). We summarize the dsSST here. 
Again, to avoid convoluting the main idea of this paper with technical details, the following description will be kept light, and the interested reader can find rigorous mathematical details in \cite{Daubechies_Lu_Wu:2011,Chen_Cheng_Wu:2014, lin2016waveshape}. {Readers familiar with speech processing will recognize a cepstral device which is also employed when calculating mel-frequency cepstral coefficients for audio signals. \cite{rabiner2011theory}}

We start by discussing how the well-known short-time Fourier transform (STFT) behaves on a given signal $f$ satisfying (\ref{Model:equation}). In general, the STFT aims to capture how the signal oscillates at different times by truncating the signal into pieces. Specifically, with a chosen window function $h$, such as a Gaussian function centered at the origin, the STFT is defined as
\begin{equation}
V^{(h)}_f(t, \xi) = \int f(\tau) h(\tau-t)e^{-i2\pi \xi (\tau-t)} \ud \tau\,,\label{eq: stft1}
\end{equation}
where $t\in\mathbb{R}$ indicates time and $\xi\in\mathbb{R}$ indicates frequency. We call $|V^{(h)}_f(t, \cdot)|^2$ the spectrogram of the signal $f$ at time $t$, since it represents the power spectrum of the truncated signal $f(\cdot) h(\cdot-t)$ around $t$. 

The motivation of the dsSST is the following fact. Due to the non-sinusoidal oscillation, at each time $t$, we will see dominant peaks around the fundamental frequency and its multiples in $|V^{(h)}_f(t, \cdot)|^2$. When there are multiple oscillatory components in $f$, multiples of the fundamental frequency of one component may mask the fundamental frequency of another component. Moreover, it is often difficult to determine whether a ridge in the spectrogram corresponds to a multiple or a fundamental tone.  We thus want to find a way to filter out these multiples while preserving the fundamental frequency of each component.
The dsSST contains as key ingredients two nonlinear operators for this purpose. These two operators are based on exploring the information hidden in the STFT itself, namely the symmetry structure between phase and frequency, and the phase information.

The first nonlinear operator takes into account the symmetry structure between phase and frequency in order to decouple the dynamical information in which we have interest (such as the IF of each component) from the artifacts (multiples) which arise when the wave-shape functions are non-sinusoidal. To achieve this goal, note that when the wave-shape function of a component is non-sinusoidal, there is an oscillatory pattern in the power spectrum $|V^{(h)}_f(t, \cdot)|^2$ whose cycles show at the fundamental frequency and its multiples. 
A naive idea which follows is that the frequency of this oscillation provides information about the period of the signal $f$. We take into account the old cepstrum idea in signal processing \cite{oppenheim2004frequency} and derive the {\em short-time cepstral transform (STCT)} for use in our dynamical setup. The STCT is defined by
\begin{equation}
C^{(h,\gamma)}_f(t, q) := \int |V^{(h)}_f(t, \xi)|^\gamma e^{-i2\pi q \xi} \ud \xi,
\label{eq: rceps1}
\end{equation}
where $\gamma>0$ is sufficiently small and $q\in\mathbb{R}$ is called the quefrency (its unit is seconds or any feasible unit in the time domain). The reason for taking the $\gamma$\textsuperscript{th} power of $|V^{(h)}_f(t, \xi)|$ is delicate. 
While $|V^{(h)}_f(t, \xi)|^2$ does oscillate, the amplitude of this oscillation changes from one cycle to another. To remove the influence of this amplitude modulation, we could take the natural logarithm of $|V^{(h)}_f(t, \xi)|$ so that the amplitude modulation is decoupled as a ``low-frequency component.'' (The remaining signal consists of cycles of constant amplitude.) However, taking the natural logarithm might be unstable numerically, so we use the approximation $ |V^{(h)}_f(t, \xi)|^\gamma$, called the ``soft logarithm.'' Systematic exploration in this regard can be found in \cite{lin2016waveshape}.
Ultimately, we obtain the fundamental {\em period} and its multiples in $C^{(h,\gamma)}_f(t, \cdot)$.

The nonlinear mask for the spectrogram is given by
\begin{equation}
U_f^{(h,\gamma)}(t,\xi):=C_f^{(h,\gamma)}(t,1/\xi),\label{Definition:NonlinearMask}
\end{equation}
where $\xi>0$ is given in $\mathrm{Hz}$. This nonlinear mask is designed by taking the fact that the two main quantities describing oscillation, namely period and frequency, are inverse to one another. Since $C^{(h,\gamma)}_f(t, \cdot)$ captures the fundamental period and its multiples at time $t$, $U_f^{(h,\gamma)}(t,\cdot)$ captures the fundamental frequency and its divisions.
Since the common information in $V^{(h)}_f(t,\cdot)$ and $U^{(h,\gamma)}_f(t, \cdot)$ is now the fundamental frequency, we can remove multiples from the STFT by
\begin{equation}
\label{eq:W}
W^{(h,\gamma)}_f(t, \xi) := V^{(h)}_f(t,\xi)U^{(h,\gamma)}_f(t, \xi),
\end{equation}
where $\xi>0$ is interpreted as frequency. The final TF representation is $|W^{(h,\gamma)}_f(t, \xi)|^2$, which is a nonlinearly filtered spectrogram. This step could be viewed as applying a nonlinear filter to the signal to remove the influence of the wave-shape function.

The second nonlinear operator takes the phase information in the STFT into account to further sharpen the nonlinearly filtered spectrogram. This nonlinear operator is produced by applying the synchrosqueezing transform \cite{Daubechies_Lu_Wu:2011,Chen_Cheng_Wu:2014}, namely
\begin{equation}\label{definition:SSTW}
SW^{(h,\gamma,\upsilon)}_{f}(t,\xi)=\int |W^{(h,\gamma)}_f(t,\eta)|^2 \delta_{|\xi-\Omega^{(h,\upsilon)}_f(t,\eta)|}\ud \eta\,,
\end{equation}
where $\xi\geq0$ and $\delta$ means Dirac measure, and the \textit{reassignment rule} $\Omega^{(h,\upsilon)}_f$ is determined by
\begin{equation}
\Omega^{(h,\upsilon)}_f(t,\xi):=
\left\{
\begin{array}{ll}
-\Im\frac{V_f^{(\mathcal{D}h)}(t,\xi)}{2\pi V_f^{(h)}(t,\xi)}&\mbox{ when }|V_f^{(h)}(t,\xi)|> \upsilon\\
-\infty&\mbox{ when }|V_f^{(h)}(t,\xi)|\leq \upsilon.
\end{array}
\right.
\,\label{RM:omega}
\end{equation}
Here, $\mathcal{D}h(t)$ is the derivative of the chosen window function $h$, $\Im$ means the imaginary part, and $\upsilon>0$ gives a threshold so as to avoid instability in the computation when $|V^{(h)}_f(t,\xi)|$ is small. $SW^{(h,\gamma,\upsilon)}_{f}$ is what we call the dsSST, and $SW^{(h,\gamma,\upsilon)}_{f}(t,\cdot)$ provides a sharpened spectrogram of the oscillatory signal at time $t$ that is free of the influence of the wave-shape function. 

\subsection{Discrete Case}

The continuous signal $f$ is uniformly sampled over a discrete set of time points with sampling interval $\Delta_t>0$. The sampling rate is hence $f_s = \Delta_t^{-1}$. Suppose the recording starts at time $t = 0$.  
Write the uniformly sampled signal as a column vector $\mathbf{f}  \in \mathbb{R}^N$, where $N$ is the number of samples and the $\ell$-th entry of $\mathbf{f}$ is $f(\ell\Delta_t)$. We index our vectors and matrices beginning with $1$. 
Choose a discrete window function $\mathbf{h} \in \mathbb{R}^{2K+1}$, a discretization of a chosen window $h$, which satisfies $\mathbf{h}(K+1) = 1$. We say that $2K + 1$ is the window length. Write $\mathbf{h}' \in \mathbb{R}^{2K+1}$ for the discretization of the derivative of the window function. For example, a discrete Gaussian window (and its derivative) with standard deviation $\sigma>0$ sampled over the interval $[-0.5, 0.5]$ at a sampling interval of $\frac{1}{2K}$ could be
\begin{gather}
\mathbf{h}(k) = e^{\frac{-\left(\frac{k-1}{2K} - 0.5 \right)^2}{2\sigma^2}}\\
\mathbf{h}'(k) = -\left(\frac{k-1}{2K} - 0.5 \right)\frac{\mathbf{h}(k)}{\sigma^2},
\end{gather}
where $k = 1, \ldots, 2K+1$.
Introduce a parameter $M$ so that $2M$ is the chosen number of pixels in the frequency axis of our time-frequency representation. 
Evaluate the STFT of $\mathbf{f}$, a matrix $\mathbf{V}_\mathbf{f} \in \mathbb{C}^{N \times 2M}$ whose entries are %
\begin{equation}
\mathbf{V}_\mathbf{f}(n, m) = \sum_{k=1}^{2K+1} \mathbf{f}(n + k - K - 1) \mathbf{h}(k) e^{\frac{-i2\pi (k-1)(m - 1)}{2M}},
\end{equation}
where $\mathbf{f}(l) := 0$ when $l < 1$ or $l > N$, $n=1,\ldots,N$ is the time index. and $m=1,\ldots,2M$ is the frequency index. 
Next, evaluate the STCT of $\mathbf{f}$. The STCT is a matrix $\mathbf{C}_\mathbf{f} \in \mathbb{C}^{N\times 2M}$ whose entries are
\begin{equation}
\mathbf{C}_\mathbf{f}(n, m') = \sum_{m=1}^{2M} \vert \mathbf{V}_\mathbf{f}(n, m) \vert^\gamma e^{\frac{-i 2\pi (m-1)(m'-1)}{2M}},
\end{equation}
where $\gamma>0$ is the chosen power parameter and $m'=1,\ldots,2M$ is the quefrency index.  We crop $\mathbf{C}_\mathbf{f}$ and consider only the first $M + 1$ columns associated with the positive quefrency axis. The inverted STCT of $\mathbf{f}$ is a matrix $\mathbf{U}_\mathbf{f} \in \mathbb{R}^{N \times (M+1)}$. For each time index $n$, consider the function $g_n \colon [0, \infty] \rightarrow \mathbf{R}$ whose known values are
\begin{gather}
g_n\left( \frac{1}{m-1} \right) = \mathbf{C}_\mathbf{f}(n, m) \quad m = 1, ..., M+1.
\end{gather}
The entries of the inverted STCT are calculated by interpolation:
\begin{gather}
\mathbf{U}_\mathbf{f}(n, m) = g_n\left( \frac{m-1}{2M} \right).
\end{gather}
The de-shape STFT of $\mathbf{f}$, a matrix $\mathbf{W}_\mathbf{f} \in \mathbb{C}^{N \times (M+1)}$, is given by the pointwise product
\begin{equation}
\mathbf{W}_\mathbf{f}(n, m) = \mathbf{V}_\mathbf{f}(n,m) \mathbf{U}_\mathbf{f}(n, m)\,,
\end{equation}
where $n=1,\ldots,N$ and $m=1,\ldots, M+1$.
To sharpen $\mathbf{W}_\mathbf{f}$, we first calculate 
\begin{equation}
\mathbf{V}'_\mathbf{f}(n, m) := \sum_{k=1}^{2K+1} \mathbf{f}(n + k - K - 1) \mathbf{h}'(k) e^{\frac{-i2\pi (k-1)(m - 1)}{2M}}.
\end{equation}
We choose a threshold $\upsilon > 0$ and calculate the reassignment operator
\begin{gather}
\bm{\Omega}_{\mathbf{f}}^{\upsilon}(n, m) = \left\{
\begin{array}{ll}
-\Im\frac{\mathbf{V}_\mathbf{f}'(n,m)}{\mathbf{V}_\mathbf{f}(n,m)} \frac{N}{2\pi(2K + 1)}&\mbox{ when }|\mathbf{V}_\mathbf{f}(n,m)|> \upsilon\\
-\infty&\mbox{ when }|\mathbf{V}_\mathbf{f}(n,m)|\leq \upsilon.
\end{array}
\right.
\end{gather}
The dsSST of $\mathbf{f}$, a matrix $S\mathbf{W}_\mathbf{f} ^\upsilon\in \mathbb{C}^{N \times (M+1)}$, is  finally given by the formula
\begin{gather}
S\mathbf{W}_\mathbf{f}^\upsilon(n, m) = \sum_{l;\, m= l-\bm{\Omega}_{\mathbf{f}}^{\upsilon}(n, m)} \mathbf{W}_\mathbf{f}(n, l).
\end{gather}
We could also use the reassignment operator to obtain the SST of $\mathbf{f}$, denoted as $S\mathbf{V}_\mathbf{f}^\upsilon\in \mathbb{C}^{N \times (M+1)}$, using the formula
\begin{gather}
S\mathbf{V}_\mathbf{f}^\upsilon(n, m) = \sum_{l;\, m= l- \bm{\Omega}_{\mathbf{f}}^{\upsilon}(n, m)} \mathbf{V}_\mathbf{f}(n, l) .
\end{gather}

\subsection{Numerical implementation} We provide numerical details for both the discrete dsSST and the discrete SST. First, to decrease the computational load, when suitable, we suggest to evaluate the STFT of $\mathbf{f}\in\mathbb{R}^N$ at only a subset of its $N$ sampling points.  For example, choosing a ``hop'' of $s\in\mathbb{N}$ samples means we calculate only those rows $\mathbf{V}_\mathbf{f}(n, \cdot)$ for which $n = s, 2s, \ldots, \lfloor \frac{N}{s} \rfloor s$. Here, $\lfloor x \rfloor$ means the largest integer smaller than $x>0$.  Second, we take the real part of the STCT and lift negative entries to zero.  Third, we select a least upper bound $u>0$ on the fundamental frequency of all components expected in $f$.  This bound allows us to eliminate envelope information from the STCT below the quefrency $1/u$.  When performing reassignment in both the dsSST and the SST, the threshold $\upsilon$ is selected by taking the $p$\textsuperscript{th} quantile of the values 
\begin{gather}
\left\{ \vert\mathbf{V}_\mathbf{f} (n, m)\vert : n = s, 2s,..., \left\lfloor \frac{N}{s} \right\rfloor s, \ m = 1, ..., M + 1 \right\}.
\end{gather}
Selecting $p$ to be a higher quantile results in the removal of noise. Since the STFT may exhibit large changes in norm over time, the threshold is determined as a function of each time point $m = 1, ..., M+1$. 
When calculating the inverted STCT, we use shape-preserving piecewise cubic interpolation.  Finally, we suggest to remove the low-frequency information below $l>0$ from all of our time-frequency representations because it tends to dominate the visualization. For example, in this paper, based on physiological knowledge, we remove information below the frequency $0.1$ $\mathrm{Hz}$. MATLAB code for performing all of the above-mentioned time-frequency analysis techniques is available at \url{https://github.com/jrvmalik}.

\section{Recycling algorithm}

The recycling algorithm is demonstrated by extracting hemodynamic information from the cardiogenic artifact in the IP. 
Our recycling algorithm uses information solely from the IP.
There are two steps to our algorithm. First, the IHR information is extracted from the dsSST of the IP. The parameters for the dsSST are given in Table~\ref{Table:0}.
Parameters were chosen  in an \textit{ad hoc} fashion. The IHR is deemed to be the dominant curve in the dsSST exceeding $50$ beats per minute ($\mathrm{bpm}$). This dominant curve is extracted by performing the following optimization {\cite[(15)]{Chen_Cheng_Wu:2014}}:
\begin{align}\label{Eq:curve}
\texttt{curve} = \arg\max_{\mathbf{c} \in{\mathbb{N}}^{\left\lfloor \frac{N}{s} \right\rfloor}} &\left\lbrace \sum_{k=1}^{\left\lfloor \frac{N}{s} \right\rfloor} S\mathbf{W}_\mathbf{f}^\upsilon(ks, \mathbf{c}(k))- \lambda \sum_{k=2}^{\left\lfloor \frac{N}{s} \right\rfloor} [\mathbf{c}(k) - \mathbf{c}(k-1)]^2 \right\rbrace\,.
\end{align}
The dominant curve $\texttt{curve}$ is a function of a subset of the sampling points of $\mathbf{f}$ and assumes positive integer values between $1$ and $M + 1$. The second term in the objective functional penalizes the curve for performing large jumps in the frequency axis between two consecutive time points; that is, it controls the regularity of the extracted heart rate. 
Our implementation relies on a choice of $\lambda = 1$. 
Applying a linear transformation using the sampling rate $f_s$ of $f$ yields an estimate $\widehat{\mathsf{IHR}}$ for the IHR in beats-per-minute ($\mathrm{bpm}$):
\begin{equation}
\widehat{\mathsf{IHR}}(k) := \frac{60 \times f_s(\texttt{curve}(k) - 1)}{2M}\,,
\end{equation}
where $k = 1, 2, \ldots, \left\lfloor \frac{N}{s} \right\rfloor$. {In our numerical experiments, the search for the dominant curve includes the following additional steps. First, perform time-averaging of the deshape matrix to obtain a power spectrum-like signal 
\begin{gather}
P_{\mathbf{f}}(m) := \sum_{k = 1}^{\left\lfloor \frac{N}{s} \right\rfloor} \left\vert S\mathbf{W}_{\mathbf{f}}^\upsilon(ks, m) \right\vert^2,\label{22}
\end{gather}
where $m = 1, ..., M + 1$. 
Next, detect the maximum
\begin{gather}
\tau = \arg \max \left\lbrace P_{\mathbf{f}}(m) : 50 \leq \frac{60 \times f_s(m - 1)}{2M}\right\rbrace.
\end{gather}
The figure $\tau$ plays the role of an initial heart rate range estimator. 
Finally, optimize \eqref{Eq:curve} while ensuring that the optimizer $\mathbf{c}\in{\mathbb{N}}^{\left\lfloor \frac{N}{s} \right\rfloor}$ satisfies the constraint
\begin{gather}
50 \leq \frac{60 \times f_s(\mathbf{c}(k) - 1)}{2M} \leq \frac{60 \times f_s(\tau - 1)}{2M} + 30\label{24}
\end{gather}
for all $k = 1, ..., \left\lfloor \frac{N}{s} \right\rfloor$.  In other words, the extracted curve may exceed the initial heart rate estimate by at most $30$ $\mathrm{bpm}$. 
This extra step reduces the risk of detecting a non-vanishing multiple of the cardiac frequency when no human input is provided into the curve extraction process. Indeed, as is discussed in \cite{lin2016waveshape}, harmonics may sometimes not be completely eliminated, or artifacts may arise due to issues like a missing fundamental tone, stacked harmonics, or numerical instability when inverting the discrete STCT.
}

We evaluate the effectiveness of the IHR estimation in the following way. Using the accompanying electrocardiogram, the ground-truth IHR, which is viewed as a continuous function on $\mathbb{R}$, is sampled at $r_i$ with the value
$\frac{1}{r_i-r_{i-1}}$,
where $r_i$ is the location (in seconds) of the $i^\text{th}$ $\mathrm{R}$ peak in the electrocardiogram. Suppose there are $N_R$ detected R peaks.  Using shape-preserving piecewise cubic interpolation over $\{r_i, \frac{1}{r_i-r_{i-1}}\}_{i=1}^{N_R}$, we recover the ground-truth IHR series, $\mathsf{IHR}\in \mathbb{R}^{\left\lfloor \frac{N}{s} \right\rfloor}$, where the $k$-th entry of $\mathsf{IHR}$ is the ground-truth IHR at time $\frac{ks}{f_s}$ in $\mathrm{bpm}$.
The following root mean square error (RMSE) metric is used to compare the estimated heart rate, $\widehat{\mathsf{IHR}}$, to the ground-truth heart rate, $\mathsf{IHR}$.
\begin{gather}
\mathrm{RMSE} = \sqrt{\frac{1}{\left\lfloor \frac{N}{s} \right\rfloor}\sum_{k=1}^{\left\lfloor \frac{N}{s} \right\rfloor} \left[ \mathsf{IHR}(k) - \widehat{\mathsf{IHR}}(k) \right]^2}.
\end{gather}
Note that these values appear in units of beats-per-minute ($\mathrm{bpm}$).
In clinical practice, physicians make decisions based on the {\em ongoing heart rate}, which is an average over a maximum delay of $10$ seconds of the IHR, according to the ANSI/AAMI standard \cite{AMS:2007}. 
For this clinical need, we consider $\mathsf{IHR}_{10}$, which is the signal obtained after applying a 10-second bi-directional moving average filter to $\mathsf{IHR}$, and we evaluate the 10-second RMSE, defined as
\begin{gather}
\mathrm{RMSE10} = \sqrt{\frac{1}{\left\lfloor \frac{N}{s} \right\rfloor}\sum_{k=1}^{\left\lfloor \frac{N}{s} \right\rfloor} \left[\mathsf{IHR}_{10}(k) - \widehat{\mathsf{IHR}}(k) \right]^2}\,.
\end{gather}

\begin{table*}
\centering
\scriptsize
\caption{Parameters for the SST, the dsSST, and curve extraction}\label{Table:0}
\begin{tabular}{lll}
\hline\noalign{\smallskip}
Parameter & Value ($\mathrm{IHR}$ extraction) & Value (source separation)\\
\noalign{\smallskip}\hline\noalign{\smallskip}
Input Sampling Rate $f_s$ & $64$ $\mathrm{Hz}$ & $64$ $\mathrm{Hz}$ \\
Window Length $2K + 1$ & $4001$ samples & $4001$ samples\\
Window Type & Gaussian & Gaussian\\
Frequency Resolution $M$ & $15000$ & $15000$ \\
Reassignment Quantile $p$ & $60$ & $0$\\
Hop $s$ & $16$ samples & $1$ sample\\
Power $\gamma$ & $0.3$ & N/A\\
Lower Bound $l$ & {$\frac{5}{6}$ $\mathrm{Hz}$} & $0.1$ $\mathrm{Hz}$\\
Upper Bound $u$& {$4.0$ $\mathrm{Hz}$} & $2.0$ $\mathrm{Hz}$\\
Heart Rate Search Range & {$50$-$240$ $\mathrm{bpm}$} & N/A\\
Curve Penalty $\lambda$ & 1 & N/A\\
\noalign{\smallskip}\hline
\end{tabular}\\
\end{table*}

The second step in our algorithm is the separation of the cardiac and respiratory signals. The respiratory signal is isolated using an approach which may be viewed as applying an {\em adaptive time-varying bandpass filter}, which is nonlinear in nature. The theoretical foundation has been established in \cite[Theorem 3.3]{Daubechies_Lu_Wu:2011}, and the technique was used previously in several places, for example, \cite{Lin_Wu_Hsu_Wang_Huang_Huang_Lo:2016}. The respiratory signal $\mathbf{r} \in \mathbb{R}^{\left\lfloor \frac{N}{s} \right\rfloor}$ is reconstructed from the SST of $\mathbf{f}$ as
\begin{equation}\label{Recovery:Resp}
\mathbf{r}(k) = \mathfrak{Re}\left\{ \sum_{m;\,{0.1 \leq \frac{(m -1)f_s}{2M}\leq \frac{\widehat{\mathsf{IHR}}(k)}{60} - 0.2} } \frac{S\mathbf{V}_\mathbf{f}^{\upsilon} (ks, m)}{M}\right\}\,,
\end{equation}
where $k = 1,2, \ldots, \left\lfloor \frac{N}{s} \right\rfloor$ and $\mathfrak{Re}$ denotes taking the real part. Removing the respiratory signal $\mathbf{r}$ from the IP $\mathbf{f}$ yields an approximation of the cardiac signal. Since the cardiac signal may contain unwanted low-frequency information coming from the spectral linkage of the respiratory signal and noise, we apply a bi-directional highpass third-order Butterworth filter with cutoff frequency $0.5$ $\mathrm{Hz}$ as a post-processing step, and we add the low-frequency component back to the separated respiratory signal.

\subsection{Comparison with other approaches}\label{Sect:compare}
To achieve a fair comparison, we compare our algorithm with other existing methods. While we are the first to focus on reconstructing the cardiac signal from the IP signal, other algorithms have indirectly achieved a reconstruction of the cardiac signal in their attempts to clarify the respiratory signal.  In \cite{Poupard2008}, the cardiogenic artifact is extracted from the thoracic impedance signal using a smoothing cubic spline filter. We implement the same smoothing cubic spline filter with a cutoff of $0.5$ Hz. Since the ground-truth cardiac activity present in the IP recording is difficult to assess, only a qualitative comparison of the two algorithms is provided. 

We may, however, augment the filtering in \cite{Poupard2008} with one of two spectral methods to achieve an estimate for the IHR which rivals the dsSST approach. To show that the advantages of the dsSST include a cancellation of respiratory harmonics exceeding the threshold $0.5$ Hz, we do the following. First, remove the low-frequency component of the IP  and obtain the signal $\mathbf{f}_{\text{hi}}$. Second, estimate the IHR by either:  extracting the dominant curve from the SST $S\mathbf{V}_{\mathbf{f}_{\text{hi}}}^\upsilon$ in the range $50$-$240$ bpm, or determining the highest peak in the power spectrum (DFT) of $\mathbf{f}_{\text{hi}}$ in the range $50$-$240$ bpm. {The curve extraction method applied to the SST matrix is the same as the curve extraction method applied to the dsSST matrix.}
To evaluate whether the proposed recycling algorithm surpasses these two additional estimates for the IHR, we apply a one-sided Wilcoxon signed-rank test under the null hypothesis that the difference between the pairs follows a symmetric distribution around zero. 
We consider the significance level of {$0.05$}.

\section{Result}

We retrospectively analyze a  data set containing IP signals from a prospective, randomized study conducted at the tertiary medical center, Chang Gung Memorial Hospital, in Linkou, New Taipei, Taiwan. The study protocol was approved by the Chang Gung Medical Foundation Institutional Review Board (No.104-0872C). All of the enrolled patients provided written informed consent. 
IP signals were recorded using the Philips Patient Monitor MP60 from subjects undergoing procedural sedation during bronchoscopy examinations for the sake of pulmonary disease diagnosis. 
The IP signals were recorded at a sampling rate of 64 $\mathrm{Hz}$.
Thirty-five subjects were enrolled in the study. 

\begin{figure*}
\centering
 \includegraphics[width=.8\textwidth]{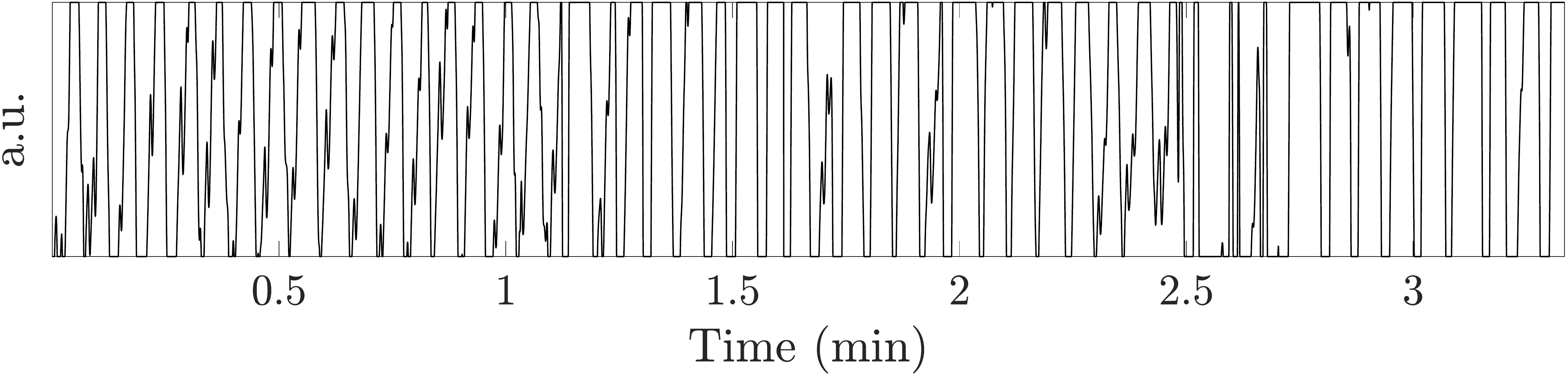}\\
 \vspace{0.1in}
 \includegraphics[width=.8\textwidth]{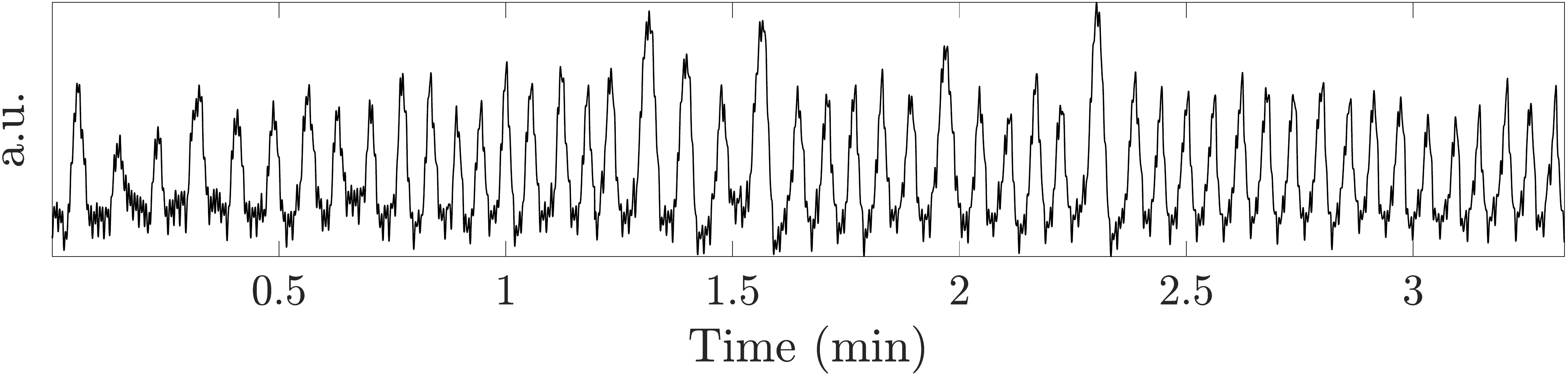}
\caption{We plot short segments of two impedance pneumographies. The first segment is out-of-range and was rejected;  the second signal was judged to be of high quality and was included in our analysis.}\label{Fig:0}
\end{figure*}

The 35 recordings were examined for over-saturation (i.e. out-of-range). We attempted to select a contiguous, in-range, $10$-minute segment from each recording. Eleven subjects were excluded from our analysis because no such segment existed. Five subjects were excluded because the corresponding electrocardiogram was of low quality. Nineteen subjects were left for our analysis.
In Fig.~\ref{Fig:0}, we show two impedance pneumographies; the first signal is a typical over-saturated signal and was rejected, and the second signal was judged to be suitable and was included in our analysis.   

\begin{figure*}
\centering
 \includegraphics[width=.8\textwidth]{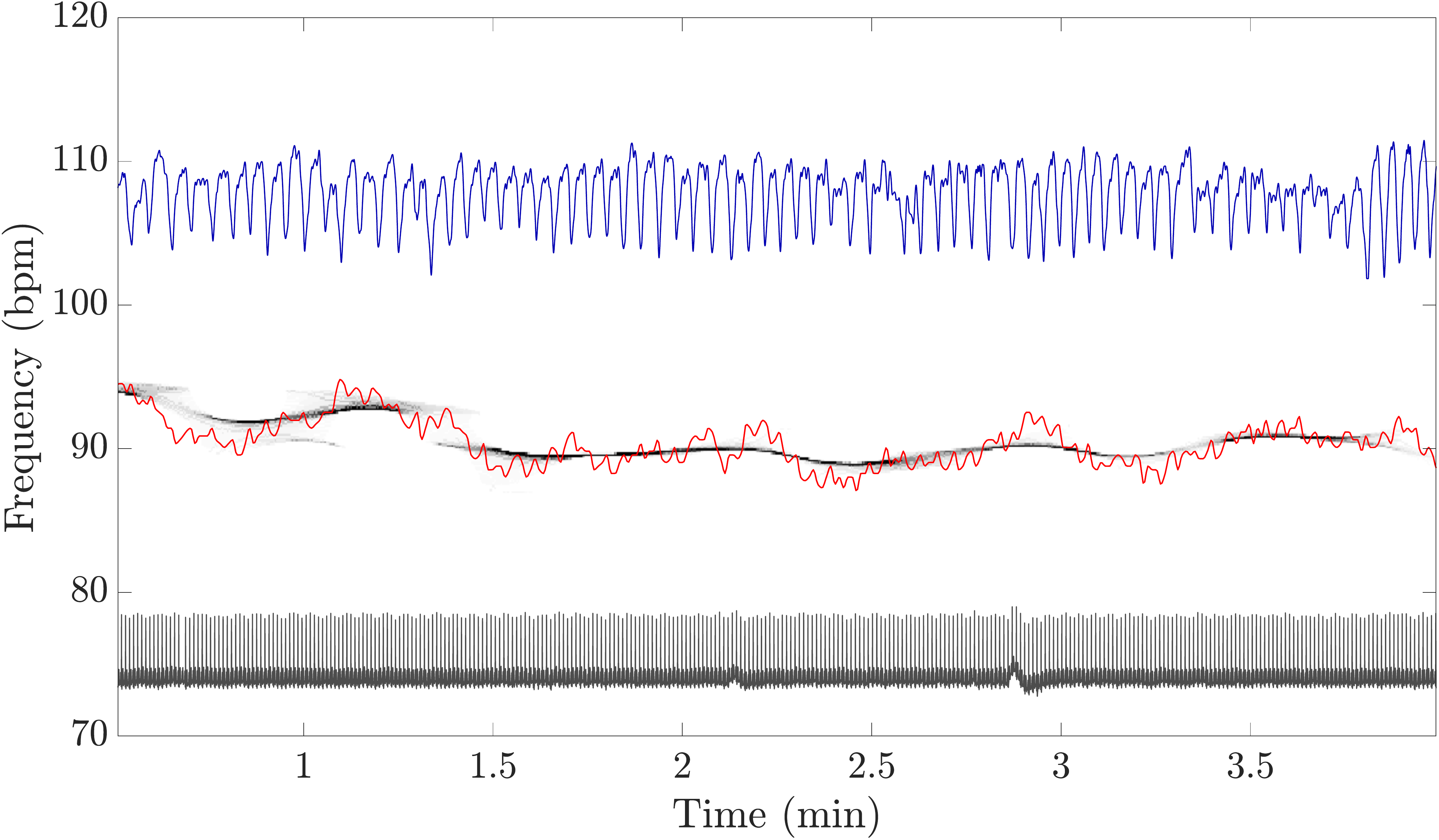}
\caption{We plot the squared modulus of the dsSST of a $3$-minute IP.  In red, we show the ground-truth IHR obtained from the corresponding electrocardiogram. The estimate for the IHR is obtained by extracting the dominant curve in the dsSST. In blue, we superimpose the original IP. In gray, we superimpose the corresponding electrocardiogram.}\label{Fig:1}
\end{figure*}

A portion of the dsSST of a $10$-minute IP is displayed in Fig.~\ref{Fig:1}.  In red, we show the ground-truth IHR obtained from the corresponding electrocardiogram. The estimate $\widehat{\textsf{IHR}}$ for the IHR is obtained by extracting the dominant curve in the dsSST. In blue, we superimpose the original IP. In gray, we superimpose the corresponding electrocardiogram.
In Table~\ref{Table:1}, we show the RMSE and RMSE10 values for the 19 segments selected for analysis. The RMSE and RMSE10 values for all subjects were {$2.29 \pm 0.74$ $\mathrm{bpm}$ (mean $\pm$ standard deviation) and $1.62 \pm 0.78$ $\mathrm{bpm}$}, respectively. 
 

\begin{table}[]
\centering
\scriptsize
\caption{Heart rate estimation error by different methods}\label{Table:1}
\begin{tabular}{lllllll}
\hline\noalign{\smallskip}
              & \multicolumn{2}{c}{dsSST}                       & \multicolumn{2}{c}{SST}                               & \multicolumn{2}{c}{HPF}                               \\
              & \multicolumn{1}{c}{RMSE} & \multicolumn{1}{c}{RMSE10} & \multicolumn{1}{c}{RMSE} & \multicolumn{1}{c}{RMSE10} & \multicolumn{1}{c}{RMSE} & \multicolumn{1}{c}{RMSE10} \\
              \noalign{\smallskip}\hline\noalign{\smallskip}
Subject 1     & 2.26                     & 1.48                       & 2.26 & \textbf{1.48} & 12.62                    & 12.49                      \\
Subject 2     & 1.69                     & \textbf{0.70}                       & 1.69                     & 0.70                       & 1.91                     & 1.09                       \\
Subject 3     & 3.11                     & \textbf{1.10}                     & 3.59                     & 2.05                       & 5.56                     & 4.70                       \\
Subject 4     & 2.41                     & \textbf{1.42}                      & 2.61                     & 1.74                       & 3.11                     & 2.39                       \\
Subject 5     & 2.59                     & \textbf{2.19}                       & 42.81                    & 42.76                      & 45.77                    & 45.72                      \\
Subject 6     & 3.13                     & 3.54                       & 1.25                     & \textbf{1.06}                       & 4.18                     & 4.01                       \\
Subject 7     & 2.03                     & 1.43                      & 2.01                     & \textbf{1.40}                       & 2.55                     & 2.07                       \\
Subject 8     & 2.27                     & \textbf{1.46}                       & 2.38                     & 1.63                       & 3.22                     & 2.71                       \\
Subject 9     & 1.78                     & \textbf{1.36}                       & 2.41                    & 2.11                      & 2.72                     & 2.44                       \\
Subject 10    & 2.02                     & \textbf{1.31}                       & 2.04                     & 1.32                       & 8.39                     & 8.22                       \\
Subject 11    & 3.77                     & \textbf{2.57}                       & 3.82                     & 2.63                       & 4.13                    & 3.03                      \\
Subject 12    & 1.16                     & \textbf{0.75}                       & 1.16                    & 0.76                       & 1.88                     & 1.63                       \\
Subject 13    & 1.77                     & {0.94}                       & 1.76                     & \textbf{0.93}                       & 4.10 & 3.79                       \\
Subject 14    & 3.73                     & {3.41}                       & 3.10                    & \textbf{2.80}                       & 26.88                    & 26.86                      \\
Subject 15    & 2.17                     & \textbf{1.70}                       & 2.18                     & 1.71                       & 2.42                     & 1.97                       \\
Subject 16    & 2.20                     & \textbf{1.54}                       & 3.25                     & 2.79                       & 4.66                     & 4.30                       \\
Subject 17    & 2.82                     & 1.68                       & 2.83                     & \textbf{1.67}                       & 3.92                     & 3.17                       \\
Subject 18    & 1.45                     & \textbf{1.26}                       & 1.46                     & 1.26                       & 62.41 & 62.40 \\
Subject 19    & 1.14                     & \textbf{0.87}                      & 1.11                     & {0.90}                       & 2.14                     & 1.95                       \\ \noalign{\smallskip}\hline\noalign{\smallskip}
Mean          & 2.29                     &\textbf{1.62}                       & 4.41                     & 3.77                       & 10.66                     & 10.26                       \\
Standard Dev. & \textbf{0.74}                     & 0.78                      & 9.08                     & 9.21                       & 16.14                    & 16.31      \\
\noalign{\smallskip}\hline
\end{tabular}
\end{table}

\begin{figure*}
\centering
 \includegraphics[width=.8\textwidth]{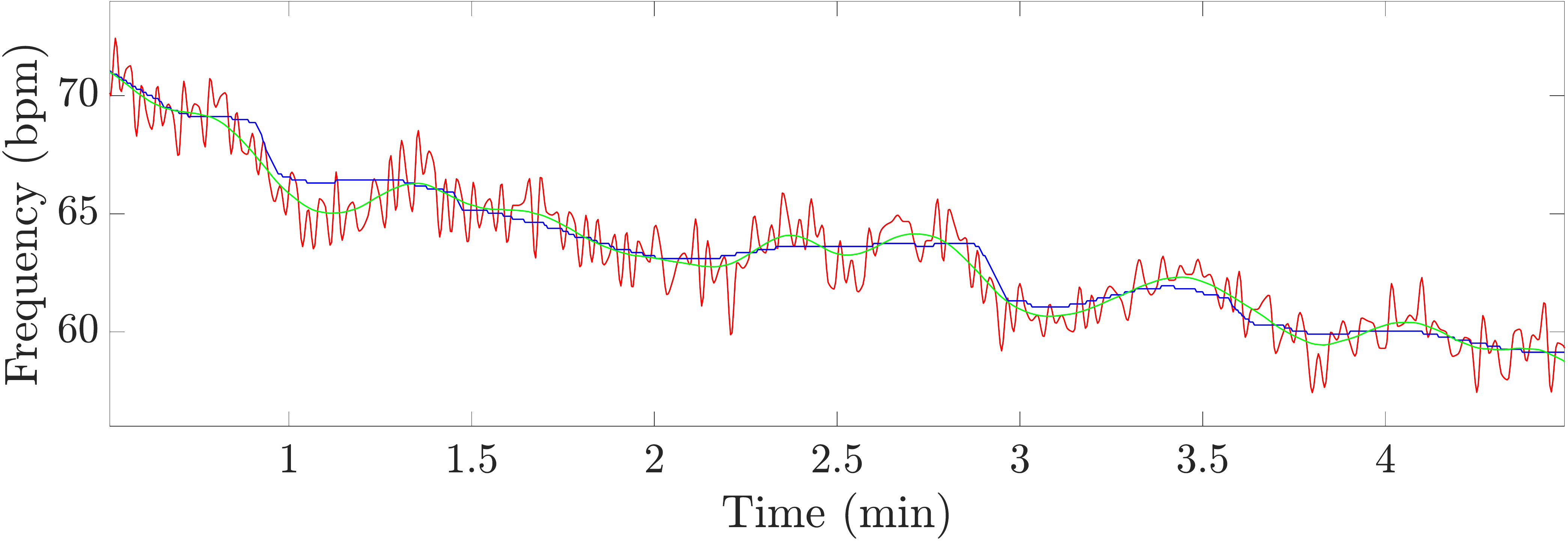}\\
\caption{We visually examine the IHR estimate afforded by the dominant curve in the dsSST. We show the ground-truth IHR determined from the electrocardiogram in red, the estimated IHR in blue, and the mean-filtered ground-truth IHR in green.}\label{Fig:2}
\end{figure*}

Discrepancies between the ground-truth IHR and the IHR estimated from the dsSST can be explained by visually inspecting both signals.  In Fig.~\ref{Fig:2}, we show the ground-truth IHR in red, the estimated IHR in blue, and the mean-filtered ground-truth IHR in green. Evidently, the estimated IHR coincides with the trend in the ground-truth IHR, and the error between the estimate and the ground-truth arises from high-frequency fluctuations in the beat-to-beat interval series.

\begin{figure*}
\centering
\includegraphics[width=.8\textwidth]{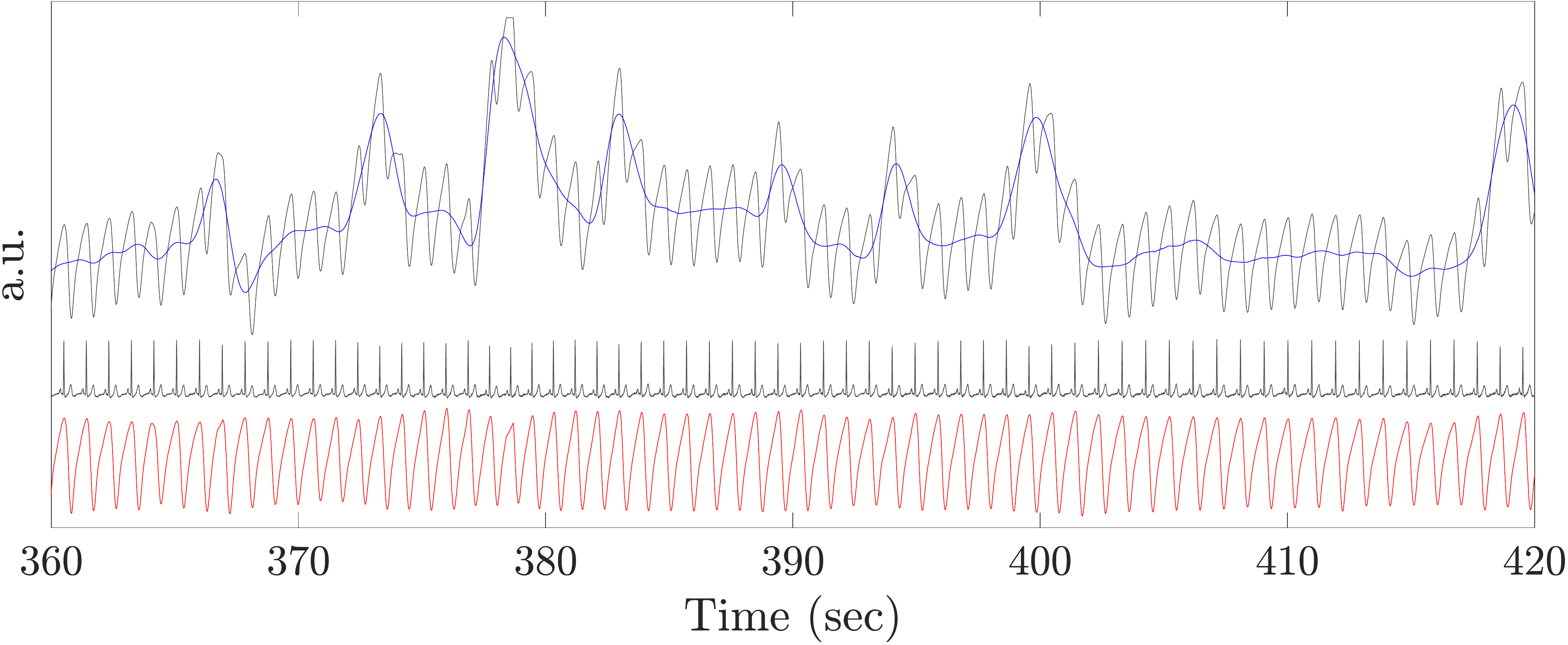}\\
\vspace{0.1in}
 \includegraphics[width=.8\textwidth]{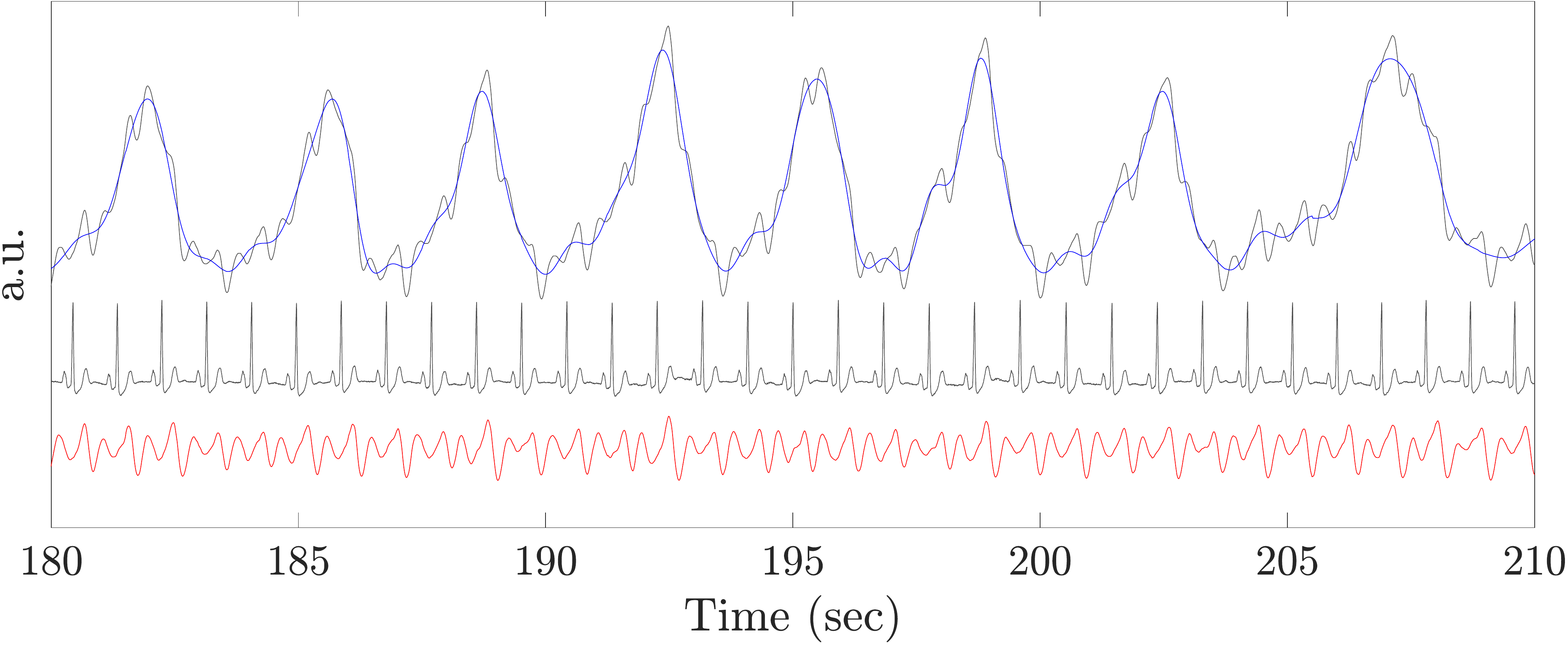}\\
\caption{We show the separation of the respiratory and cardiac signals. At the top of each plot, we show the respiratory signal in blue and the IP in gray. At the bottom of each plot, we show the extracted hemodynamic signal in red. In the middle of each plot, we show the simultaneously recorded electrocardiogram in gray. }\label{Fig:3}
\end{figure*}

In Fig.~\ref{Fig:3}, we show the result of the second step of our algorithm: separation of the cardiac and respiratory signals. At the top of each plot, we show a segment of the clarified respiratory signal in blue overlying the original IP in gray. At the bottom of each plot, we show a segment of the cardiac signal in red, aligned with the original electrocardiogram in gray. Note the correspondence between the electrocardiogram's wave-shape and the extracted cardiac signal's wave-shape. 

In Table \ref{Table:1}, we show the heart rate estimation error for the two additional methods described in \ref{Sect:compare}.  These methods were derived from the existing lowpass filter approach to clarifying the respiratory signal \cite{Poupard2008}. The heading SST indicates that the dominant curve was extracted from the time-frequency representation $S\mathbf{V}_{\mathbf{f}_{\text{hi}}}^\upsilon$, and the heading HPF indicates that the IHR was estimated by finding a local maximum in the {power spectrum} of the filtered signal $\mathbf{f}_{\text{hi}}$. It is clear that by using the SST, we can obtain a reasonable estimate for heart rate information in the majority of cases. The deshape step improves the result in the most difficult cases. Finally, the quality of the HPF estimate is not as good as the SST or dsSST estimates.  {For example, subject 1 possessed a heart rate which was steadily increasing over the duration of the recording, resulting in high $\mathrm{RMSE}$ and $\mathrm{RMSE10}$ values for the HPF method. 
The power spectrum of subject 11's filtered IP signal did not decay quickly enough, resulting in an IHR estimate close to $50$ $\mathrm{bpm}$ for the HPF method.  
The power spectrum of subject 14's filtered IP signal contained strong cardiac harmonics which overpowered the fundamental cardiac frequency. 
The synchrosqueezed spectrogram of subject 5 contained a dominant respiratory harmonic, resulting in a poor IHR estimate for both the HPF and SST estimation methods.}
Evidently, the {HPF and SST estimation methods} are {not immune to detecting} respiratory harmonics, {while the HPF method suffers when the signal is noisy and when the heart rate changes significantly during the recording.} 
The Wilcoxon signed-rank test showed a statistically significant improvement in RMSE (respectively RMSE10) by the proposed dsSST algorithm over the SST method with the $p$-value {$4.36 \times 10^{-2}$} (respectively {$3.35 \times 10^{-2}$}). {In comparison with the HPF method, the $p$-statistics for improvement in RMSE (respectively RMSE10) were both less than $10^{-4}$.}

\begin{figure*}
\centering
\includegraphics[width=.8\textwidth]{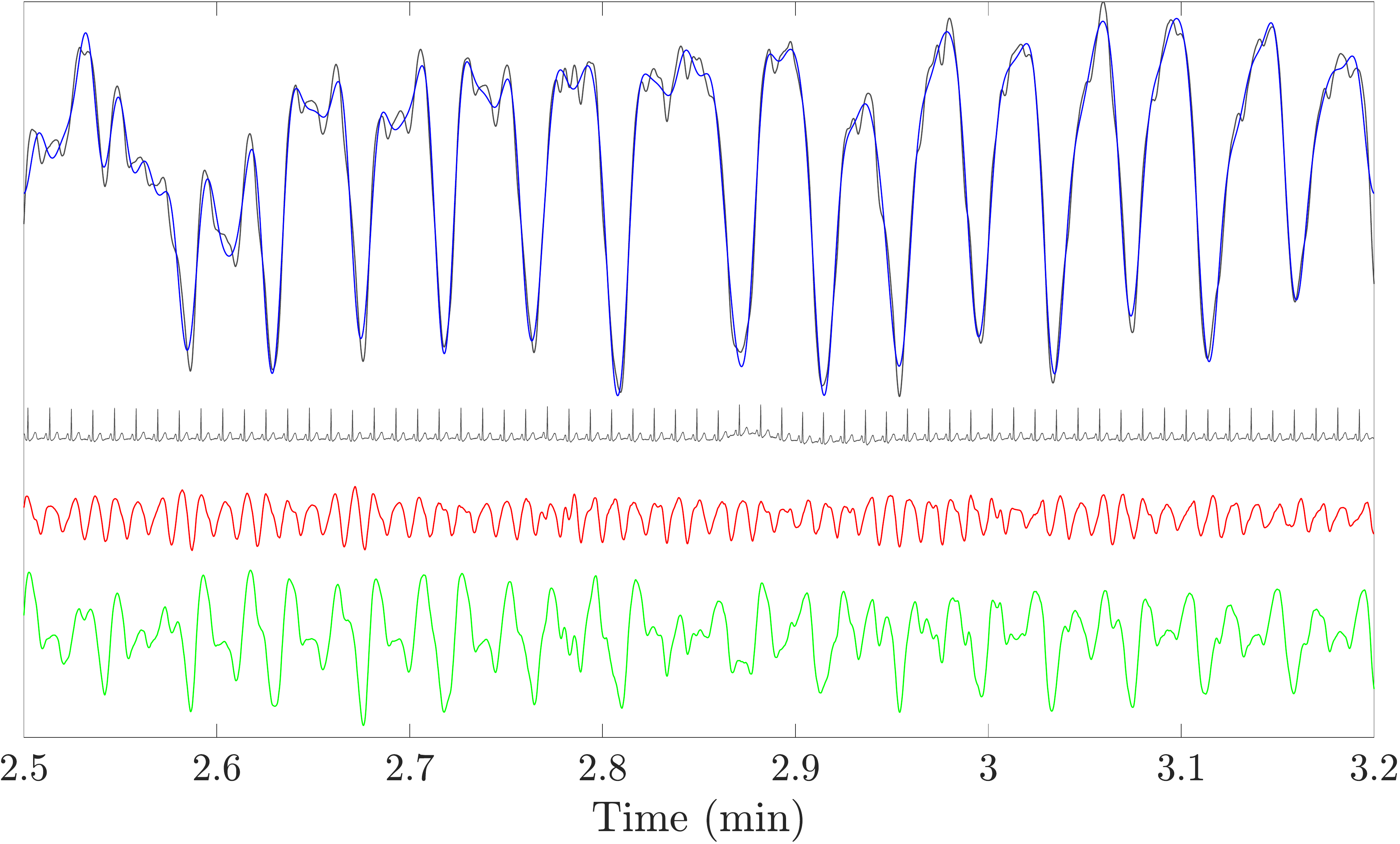}\\
\caption{We compare the separation of the respiratory and cardiac signals by different approaches. The IP is shown in gray, over which the extracted respiratory signal by the proposed recycling algorithm is shown in blue. Below the IP signal, the ECG signal is shown in gray. The extracted hemodynamic signal by the proposed recycling algorithm is shown in red, and the extracted hemodynamic signal by the high pass filter of \cite{Poupard2008} is shown in green. }\label{Fig:4}
\end{figure*}

See Figure \ref{Fig:4} for a qualitative comparison of the extracted hemodynamic signal by the proposed recycling algorithm and the highpass filter described in \cite{Poupard2008}. It is clear that the oscillatory pattern of the hemodynamic signal extracted by the highpass filter is less regular.

%

\section{Discussion and Conclusion}

We recycle the physiological information, the cardiogenic artifact, that is commonly discarded from biomedical measurements such as the IP signal. 
Our algorithm retrieves IHR information and extracts the oscillatory signal reflecting the hemodynamic activity. 
The corresponding MATLAB code is made publicly available so that our methods may be reproduced. 
{The dsSST succeeds by masking respiratory harmonics in the spectrogram of the IP signal, allowing the fundamental frequency of the cardiac component to be unambiguously estimated.
This time-varying estimate is used to guide an adaptive filtration of the IP signal, leading to a separation of the cardiac and respiratory signals. Algorithms which were previously used to remove cardiogenic artifacts cannot lead to a robust IHR estimate because they fail to disregard respiratory harmonics.}
In our demonstration, which was performed on real signals recorded from 19 subjects, the RMSE and RMSE10 values for the heart rate estimation stage were {$2.29 \pm 0.74$ $\mathrm{bpm}$ and $1.62 \pm 0.78$ $\mathrm{bpm}$}, respectively. A comparison with other approaches {confirms} the superiority of the proposed recycling algorithm. This result indicates that the proposed recycling algorithm can provide an accurate estimate for the heart rate over a 10-second time frame.
When only a limited number of sensors is available, this algorithm maximizes the amount of information which can be relayed to medical professionals.

It is worth noting that the output of our algorithm should be trusted when the original signal is properly recorded. In general, the accuracy of any algorithm degrades if the signal is not properly recorded. In the case of the IP, when the signal is out-of-range, there is no information recorded, and no cardiogenic artifact can be analyzed. Practitioners should ensure that their monitoring devices are properly set up. Additionally, to avoid the out-of-range issue, a signal quality index could indicate how much we should trust the result.

We discuss some points specific to the IP. Since the hemodynamic information extracted by the first step of our algorithm tends to align with a mean-filtered version of the ground-truth IHR signal, its suitability for HRV analysis might not be optimal. We are able to account for low-frequency variability, but high-frequency variability is typically unobserved. 
Hence, while it could provide extra information about heart rate with accuracy less than $2$ $\mathrm{bpm}$, we should use the extracted hemodynamic information as merely an auxiliary resource when doing traditional HRV analysis.

Second, although the respiratory and cardiac signals have well-separated frequencies, the multiples associated with the non-sinusoidal wave-shape function approximating the respiratory signal are not negligible. %
This is the main reason why the performance of the highpass filter is not comparable with the proposed recycling algorithm -- it is non-adaptive to the signal. For patients with higher respiratory rate, the multiples will have a stronger influence on the heart rate spectrum. The same reason explains why the SST performs worse -- the existence of multiples associated with the respiratory signal contaminates the heart rate information. 
On the other hand, when multiples associated with the IP signal are not too strong, the adaptive time-varying bandpass filter strategy in \eqref{Recovery:Resp} provides a reasonable-enough recovery of the respiratory signal, and hence the cardiac oscillation. 
Note that in this IP example, the possible clinical application of the recovered cardiac waveform needs further exploration; for example, it would be interesting to ask which hemodynamic information could be extracted by analyzing its morphology.
For other biomedical signals, particularly when the frequencies of different components are close and/or the wave-shape functions are complicated, we may need a more sophisticated approach for the separation. For example, to extract the fetal electrocardiogram from the trans-abdominal maternal electrocardiogram, 
due to the overlapping of spectra of the wave-shape functions approximating the oscillatory patterns of the fetal and maternal components, the manifold learning approach (e.g. non-local Euclidean median) is needed to achieve an efficient separation \cite{Su_Wu:2016b}. A systematic study in this regard will be reported in the near future. 

Finally, we mention that the proposed model and our algorithm has the potential to be applied to other oscillatory biomedical signals in the field of health monitoring, where the goal is to transform numerical information (hidden or not) from any of the growing number of measurement devices into a format which can be delivered to and interpreted by medical professionals. However, we also need to mention that the proposed model and algorithm do not take structured noise (in addition to the cardiogenic artifact) into account, which is the main challenge when dealing with mobile devices. Thus, to apply the proposed model and algorithm to mobile devices for mobile health, more environmental information or extra channels are needed to help deal with the inevitable noise artifacts. We will systematically explore this potential in our future work.

\section*{Acknowledgements}
The authors acknowledge Dr. Yu-Lun Lo for sharing the data.  The authors declare no conflicts of interest.

\bibliographystyle{amsplain}
\bibliography{revisionv1_arxiv}   

\providecommand{\bysame}{\leavevmode\hbox to3em{\hrulefill}\thinspace}
\providecommand{\MR}{\relax\ifhmode\unskip\space\fi MR }
\providecommand{\MRhref}[2]{%
  \href{http://www.ams.org/mathscinet-getitem?mr=#1}{#2}
}
\providecommand{\href}[2]{#2}
\begin{thebibliography}{10}

\bibitem{BrownBarber1994}
B.~Brown, D.~Barber, A.~Morice, and A.~Leathard, \emph{{Cardiac and respiratory
  related electrical impedance changes in the human thorax}}, IEEE Trans.
  Biomed. Eng. \textbf{41} (1994), 729--734.

\bibitem{Bucklar2003}
G.~B. Bucklar, V.~Kaplan, and Konrad~E. Bloch, \emph{{Signal processing
  technique for non-invasive real-time estimation of cardiac output by
  inductance cardiography (thoracocardiography)}}, Medical and Biological
  Engineering and Computing \textbf{41} (2003), no.~3, 302--309.

\bibitem{Chen_Cheng_Wu:2014}
Y.-C. Chen, M.-Y. Cheng, and H.-T. Wu, \emph{{Nonparametric and adaptive
  modeling of dynamic seasonality and trend with heteroscedastic and dependent
  errors}}, J. Roy. Stat. Soc. B \textbf{76} (2014), 651--682.

\bibitem{Daubechies_Lu_Wu:2011}
I.~Daubechies, J.~Lu, and H.-T. Wu, \emph{Synchrosqueezed wavelet transforms:
  An empirical mode decomposition-like tool}, Appl. Comput. Harmon. Anal.
  \textbf{30} (2011), 243--261.

\bibitem{Folke2003}
M.~Folke, L.~Cernerud, M.~Ekstr{\"{o}}m, and B.~H{\"{o}}k, \emph{{Critical
  review of non-invasive respiratory monitoring in medical care}}, Medical {\&}
  Biological Engineering {\&} Computing \textbf{41} (2003), no.~4, 377--83.

\bibitem{AMS:2007}
Association for the Advancement~of Medical~Instrumentation et~al.,
  \emph{American national standard: cardiac monitors, heart rate meters, and
  alarms}, ANSI/AAMI EC13:2002/(R)2007, 2007,
  http://www.pauljbennett.com/pbennett/work/ec13/ec13.pdf.

\bibitem{HouShi:2016}
T.~Y. Hou and Z.~Shi, \emph{Extracting a shape function for a signal with
  intra-wave frequency modulation}, Phil. Trans. R. Soc. A \textbf{374} (2016),
  20150194.

\bibitem{IMCStefanovska:2015}
D.~Iatsenko, P.~V.~E. McClintock, and A.~Stefanovska, \emph{Nonlinear mode
  decomposition: A noise-robust, adaptive decomposition method}, Physical
  Review E \textbf{92} (2015), 032916--1--032916--25.

\bibitem{lin2016waveshape}
C.-Y. Lin, S.~Li, and H.-T. Wu, \emph{Wave-shape function analysis--when
  cepstrum meets time-frequency analysis}, Journal of Fourier Analysis and
  Applications \textbf{24} (2018), no.~2, 451--505.

\bibitem{Lin_Wu_Hsu_Wang_Huang_Huang_Lo:2016}
Y.-Y. Lin, H.-T. Wu, C.-A. Hsu, C.-W. Wang, P.-C. Huang, Y.-H. Huang, and Y.-L.
  Lo, \emph{Sleep apnea detection based on thoracic and abdominal movement
  signals of wearable piezo-electric bands}, IEEE J. Biomed. Health Inform.
  \textbf{21} (2017), no.~6, 1533--1545.

\bibitem{capnogram:2018}
Y.-L. Lo, H.-T. Wu, Y.-T. Lin, H.-P. Kuo, and T.-Y. Lin, \emph{Hypoventilation
  patterns during bronchoscopic sedation and their clinical relevance based on
  capnographic and respiratory impedance analysis}, Journal of Clinical
  Monitoring and Computing \textbf{In press} (2019).

\bibitem{Luo1994}
S.~Luo, W.J. Tompkins, and J.G. Webster, \emph{{Cardiogenic artifact
  cancellation in apnea monitoring}}, Engineering in Medicine and Biology
  Society, 1994. Engineering Advances: New Opportunities for Biomedical
  Engineers. Proceedings of the 16th Annual International Conference of the
  IEEE, vol.~2, 1994, pp.~968--969.

\bibitem{oppenheim2004frequency}
A.~V. Oppenheim and R.~W. Schafer, \emph{From frequency to quefrency: A history
  of the cepstrum}, IEEE Signal Processing Magazine \textbf{21} (2004), no.~5,
  95--106.

\bibitem{piwek2016rise}
L.~Piwek, D.~A. Ellis, S.~Andrews, and A.~Joinson, \emph{The rise of consumer
  health wearables: promises and barriers}, PLoS Medicine \textbf{13} (2016),
  no.~2, e1001953.

\bibitem{Poupard2008}
L.~Poupard, M.~Mathieu, R.~Sart{\`{e}}ne, and M.~Goldman, \emph{{Use of
  thoracic impedance sensors to screen for sleep-disordered breathing in
  patients with cardiovascular disease}}, Physiological Measurement \textbf{29}
  (2008), no.~2, 255--267.

\bibitem{rabiner2011theory}
L.~R. Rabiner and R.~W. Schafer, \emph{Theory and applications of digital
  speech processing}, vol.~64, Pearson Upper Saddle River, NJ, 2011.

\bibitem{Sackner1991613}
M.~A. Sackner, R.~A. Hoffman, D.~Stroh, and B.~P. Krieger,
  \emph{Thoracocardiography: Part 1: Noninvasive measurement of changes in
  stroke volume comparisons to thermodilution}, Chest \textbf{99} (1991),
  no.~3, 613 -- 622.

\bibitem{Schuessler1998}
T.~F. Schuessler, S.~B. Gottfried, P.~Goldberg, R.~E. Kearney, and J.~H~T
  Bates, \emph{{An adaptive filter to reduce cardiogenic oscillations on
  esophageal pressure signals}}, Annals of Biomedical Engineering \textbf{26}
  (1998), no.~2, 260--267.

\bibitem{Seppa2011}
V.~P. Sepp{\"{a}}, J.~Hyttinen, and J.~Viik, \emph{{A method for suppressing
  cardiogenic oscillations in impedance pneumography}}, Physiological
  Measurement \textbf{32} (2011), no.~3, 337--345.

\bibitem{Smith1994}
T.~C. Smith, A.~Green, and P.~Hutton, \emph{{Recognition of cardiogenic
  artifact in pediatric capnograms}}, Journal of Clinical Monitoring
  \textbf{10} (1994), no.~4, 270--275.

\bibitem{STOREY201750}
V.~C. Storey and I.-Y. Song, \emph{Big data technologies and management: What
  conceptual modeling can do}, Data \& Knowledge Engineering \textbf{108}
  (2017), 50 -- 67.

\bibitem{Su_Wu:2016b}
L.~Su and H.-T. Wu, \emph{Extract fetal {ECG} from single-lead abdominal {ECG}
  by de-shape short time fourier transform and nonlocal median}, Frontiers in
  Applied Mathematics and Statistics \textbf{2} (2017), no.~3, 2.

\bibitem{electronics3020282}
T.~Tamura, Y.~Maeda, M.~Sekine, and M.~Yoshida, \emph{Wearable
  photoplethysmographic sensors -- past and present}, Electronics \textbf{3}
  (2014), no.~2, 282--302.

\bibitem{1741-2552-12-3-031001}
J.~A. Uriguen and B.~Garcia-Zapirain, \emph{Eeg artifact removal
  state-of-the-art and guidelines}, Journal of Neural Engineering \textbf{12}
  (2015), no.~3, 031001.

\bibitem{Weese-Mayer1989}
D.~E. Weese-Mayer, R.~T. Brouillette, A.~S. Morrow, L.~P. Conway, L.~M.
  Klemka-Walden, and C.~E. Hunt, \emph{{Assessing validity of infant monitor
  alarms with event recording}}, The Journal of Pediatrics \textbf{115} (1989),
  no.~5 PART 1, 702--708.

\bibitem{Wilson1982}
A.~J. Wilson, C.~I. Franks, and I.~L. Freeston, \emph{Methods of filtering the
  heart-beat artefact from the breathing waveform of infants obtained by
  impedance pneumography}, Medical and Biological Engineering and Computing
  \textbf{20} (1982), no.~3, 293--298.

\bibitem{Wu:2013}
H.-T. Wu, \emph{Instantaneous frequency and wave shape functions {(I)}}, Appl.
  Comput. Harmon. Anal. \textbf{35} (2013), 181--199.

\bibitem{XuYangDaubechies:2018}
J.~Xu, H.~Yang, and I.~Daubechies, \emph{Recursive diffeomorphism-based
  regression for shape functions}, SIAM Journal on Mathematical Analysis
  \textbf{50} (2018), no.~1, 5--32.

\bibitem{Yasuda2005}
Y.~Yasuda, A.~Umezu, S.~Horihata, K.~Yamamoto, R.~Miki, and S.~Koike,
  \emph{{Modified thoracic impedance plethysmography to monitor sleep apnea
  syndromes}}, Sleep Medicine \textbf{6} (2005), no.~3, 215--224.

\bibitem{Zhang2012}
Z.~Zhang, J.~Zheng, H.~Wu, W.~Wang, B.~Wang, and H.~Liu, \emph{{Development of
  a respiratory inductive plethysmography module supporting multiple sensors
  for wearable systems}}, Sensors (Switzerland) \textbf{12} (2012), no.~10,
  13167--13184.

\end{thebibliography}

\end{document}